\documentstyle[preprint,eqsecnum,aps,epsf]{revtex}
\begin{document}
% \draft command makes pacs numbers print
\draft
\preprint{
SNUTP--94--063
\hspace{-43.5mm}
\raisebox{2.5ex}{
KAIST--CHEP--95/06
}
\hspace{-40.0mm}
\raisebox{-2.5ex}{
UMN--TH--1338/95
}
}
\title{$a_1 (1260)$ contribution to  photon and dilepton \\ 
productions  from hot hadronic matter : Revisited }
\author{
Jae Kwan Kim$^{a}$,~~~~~Pyungwon Ko$^{b}$ \thanks{ pko@phyb.snu.ac.kr},
\\ Kang Young 
Lee$^{a}$  \thanks{kylee@chep5.kaist.ac.kr}, 
~~~~~Serge Rudaz$^{c}$\\  }
\address{$^{a}$  Dep. of physics, KAIST, Taejon, 305-701, Korea
\\
$^{b}$ Dep. of physics, Hong-Ik University,
 Seoul 121-791, Korea \\
$^{c}$  School of Physics and Astronomy, Univ. of Minnesota,  \\
Minneapolis, MN 55455, U.S.A.}
%\date{\today}
\maketitle
\begin{abstract}
% insert abstract here
We consider the $a_{1} (1260)$ contribution to  photon and dilepton 
productions from hot hadronic matter. The effective lagrangian with 
$\pi, \rho$ and $a_{1}$ discussed by Ko and Rudaz is employed, which 
successfully reproduces the $\Gamma (a_{1} \rightarrow \pi \gamma) = 
670$ keV.
Our results for the photon production rate are comparable to   those 
obtained by Xiong {\it et  al.}, while the dilepton production rates come
out smaller than those obtained by Gale {\it et al.}  by an order of
magnitude.
 \\
\end{abstract}
% insert suggested PACS numbers in braces on next line

% body of paper here

\narrowtext
%\tighten

\section{Introduction}
\label{sec:one}

Photon and dilepton signals have been considered as being important 
in probing quark-gluon plasma which might be formed in the initial stage of 
relativistic heavy ion collisions.  This is due to lack of the strong final 
state interactions of photons or dileptons with hadronic matter.  
Viewed as such, it is important to estimate the production rates of 
photon and dileptons from hot hadronic matter, since the latter 
may be backgrounds to the same signals from quark-gluon plasma. 

Recently, the role of $a_{1} (1260)$ resonance in  photon $(\pi \rho  
\rightarrow a_{1} \rightarrow \pi \gamma$) and dilepton ($\pi a_{1} 
\rightarrow \rho \rightarrow l^{+} l^{-}$)  
productions from hot hadronic matter was emphasized by several authors
\cite{xiong}--\cite{gale}.
These studies are based on effective chiral  lagrangians with
$\rho$ and $a_1$ with two different types. 
The coefficients in their effective lagrangians are determined by
fitting the measured $a_{1} \rho \pi$ decay rate. Then, these effective
lagrangains are further used to predict $a_{1} \rightarrow \pi \gamma$ and
$\pi a_{1} \rightarrow l^{+} l^{-}$ {\it a la} vector meson hypothesis.
Both of these works predict 
rather large a decay rate for $a_{1}^{\pm} \rightarrow \pi^{\pm} \gamma$
to be about $\sim 1.4$ MeV, compared to the experimental value, 
\begin{equation}
\Gamma(  a_{1}^{\pm} \rightarrow \pi^{\pm} \gamma ) = (0.64 \pm 0.28) ~~~
{\rm MeV},
\end{equation}
although the experimental error is quite large.
Thus, one can naively expect that photon and dilepton production  rates 
predicted by these effective lagrangians  might have been overestimated by 
factor of $\sim 2$ or so.  In calculating the dilepton production rate 
through $\pi a_{1} \rightarrow \rho \rightarrow l^{+} l^{-}$, one needs
to know the $\pi-a_{1}-\rho (\gamma)$ vertex for wide range  of
$q_{\rho}^2$. However, different effective lagrangains usually give different
$\pi-a_{1}-\rho(~{\rm or~} \gamma)$ form factors in general, and it is 
necessary to know how sensitive the results are to the specific 
lagrangians adopted.  In fact, using the effective lagrangian by Ko and 
Rudaz, we find that the dilepton production 
rate from the $\pi a_{1}$ reaction becomes smaller by an order of magnitude
compared to that obtained in Ref.~\cite{gale}, as a result of different 
behavior of $\pi-a_{1}-\rho(~{\rm or~} \gamma)$ form factors.

In this paper, we use an effective lagrangian  
considered  by Ko and Rudaz \cite{ko} to derive the $\pi-a_{1}-\rho$ vertex.  
This lagrangian was shown to successfully predict $\Gamma ( a_{1}^{\pm} 
\rightarrow \pi^{\pm} \gamma ) = 0.67$ MeV, and $\Gamma (a_{1} \rightarrow 
\rho \pi) = 328$ MeV \footnote{The decay width $\Gamma (a_{1} \rightarrow \rho 
\pi) \sim 480$ MeV obtained in Ref.~\cite{ko} was wrong unfortunately, since 
it was based on the wrong formula in Ref.~\cite{meissner}. This mistake was
pointed out in Ref.~\cite{kapusta}, but was overlooked by the authors of 
Ref.~\cite{ko}.  Results in the present work does not depend on that formula.
}.  Unlike other effective lagrangian approaches, 
the decay rate of $a_{1} \rightarrow \rho \pi$ is the prediction of 
the effective Lagrangian considered by Ko and Rudaz \cite{ko}, which
nicely compares  with the data $\sim 400$ MeV.  Therefore, it is worthwhile 
to investigate  the photon and the dilepton production rates using the
$\pi-a_{1}-\rho(~{\rm or~} \gamma)$ form factor of Ref. ~\cite{ko}.
A brief review of the model lagrangian is given in Section~\ref{sec:two}. 
Then, we use this vertex to calculate the $a_{1} (1260)$ contributions to 
photon  and dilepton productions from hot hadronic matter in  
Secs.~\ref{sec:photon} and \ref{sec:dilepton}, respectively.  
Our results are summarized and compared with others'  results  
in Sec.~\ref{sec:conclusion}.  

\section{Model lagrangian}
\label{sec:two}

The $\pi-\rho-a_1$ system is important in understanding spontaneous 
chiral symmetry breaking and its restoration. 
Although $\rho$ and $a_1$ form a parity doublet, their masses are not
degenerate because of spontaneous chiral symmetry breaking and the 
KSFR relation, which is encoded in Weinberg's sum rule. 
One can study this system using $SU(2)_{L} \times SU(2)_R$ effective chiral
lagrangian with $\rho$ and $a_1$ being treated as gauge fields associated 
with local chiral symmetry.  This hypothetical local symmetry is broken
into global symmetry via masslike terms for $\rho$ and $a_1$ fields.  
Spontaneous chiral symmetry breaking can be included either by making 
pion fields transform nonlinearly under chiral group, or by use of the 
linear sigma model with a symmetry breaking potential. 
In Ref.~[4], a gauged linear sigma model was constructed with three 
different types of masslike terms for $\rho$ and $a_1$ fields.  
In this model, $\rho$ and $a_1$ can get their masses entirely from the 
vacuum expectation value of $\sigma$ field.  
For a certain set of parameters, a successful phenomenology for $a_1$ decays
was obtained [4].  Relegating the various aspects  
of this effective lagrangian to Ref.~[4], we give a short review of the 
interaction lagrangians which are relevant to this work.

The $a_{1} \rho \pi$ vertex is given by the following interaction lagrangian,
\begin{eqnarray}
{\cal L} (a_{1} \rho \pi) & = & {g^{2} f_{\pi} \over Z_{\pi}}~\left[ 2 c 
\vec{\pi} \cdot ( \vec{\rho}_{\mu} \times \vec{a}^{\mu} ) + 
{1 \over 2 m_{a_1}^2}~ \vec{\pi} \cdot ( \vec{\rho}_{\mu\nu} 
\times \vec{a}^{\mu\nu} )      \right.
\nonumber    \\
& + & \left. \left( { 1\over m_{a_1}^2} - {\kappa_{6} Z_{\pi} 
\over m_{\rho}^2} \right) \partial_{\mu} \vec{\pi} \cdot ( 
\vec{\rho}^{\mu \nu} \times \vec{a}^{\nu} )  \right]. 
\end{eqnarray}
Using this lagrangian, one can derive the amplitude for $a_{1} (k,e)
\rightarrow \rho (k^{'},e^{'}) \pi(p)$ :  
\begin{eqnarray}
{\cal M} (a_{1} \rightarrow \rho \pi) & = & 
(T_{1})_{\mu\nu} e^{\mu} e^{' \nu} 
\label{eq:t1}
\\
&= & f_{a_{1} \rho \pi} e
\cdot e^{'} + g_{a_{1}\rho\pi} k^{'} \cdot e~ k \cdot e^{'} 
+  h_{a_{1}\rho\pi} k^{'} \cdot e k^{'} \cdot e^{'},
\label{eq:amparhopi}
\\
f_{a_{1}\rho\pi} & = & {g^{2} f_{\pi} \over Z_{\pi}}~\left[ 
2 c + {k^{'2} \over m_{a_1}^2}\right] + {\kappa_{6} g^{2} f_{\pi} \over 
m_{\rho}^2}~p \cdot k^{'},
\label{eq:farhopi}
\\
g_{a_{1} \rho \pi} & = & - {\kappa_{6} g^{2} f_{\pi} \over 
m_{\rho}^2},
\label{eq:garhopi}
\\
h_{a_{1}\rho\pi} & = & - {g^{2} f_{\pi} \over Z_{\pi}}~\left( {1 \over
m_{a_1}^2} - { \kappa_{6} Z_{\pi} \over m_{\rho}^2} \right).
\label{eq:harhopi}
\end{eqnarray}
Here, the $\rho$ and the $a_1$ mesons can be either on or off the 
mass-shell.  The last form factor $h_{a_{1}\rho \pi}$ can be  relevant  
to  a virtual $\rho$ meson such as in the process $\pi a_{1} \rightarrow
\rho \rightarrow l^+ l^-$ in principle. However, it gives a null  
result because of 
\[
\bar{v}(l_{1}) \gamma_{\mu} k^{'\mu}  u(l_{2}) = 
\bar{v}(l_{1}) \gamma_{\mu} (l_{1}^{\mu} + l_{2}^{\mu} ) u(l_{2}) = 0.
\]
Therefore, only $f_{a_{1}\rho\pi}$ and $g_{a_{1}\rho\pi}$ form factors are 
relevant to the present study.  

The above  amplitude {\it predicts} $\Gamma (a_{1} \rightarrow \rho
\pi) = 328$ MeV for $m_{a_1} = 1260$ MeV, to be compared with the PDG 
value $\sim 400$ MeV.  (We use the following numerical values : 
$g = 5.04, c = -0.12, Z_{\pi} = 0.17, \kappa_{6} = 1.25$.)
[In contrast, the decay rate for $a_{1} \rightarrow \rho \pi$ is an input 
in other approaches, in contrast with our approach.]

In case that the $\rho$ appears as a physical photon ($\gamma$), the 
above interaction lagrangian can be written as 
\begin{equation}
{\cal L} (a_{1} \pi \gamma) = {\kappa_{6} e g f_{\pi} \over m_{\rho}^2}~
B_{\mu\nu} (\vec{a}^{\mu\nu} \times \vec{\pi})_{3},
\end{equation}
for real or virtual $a_{1}$.  
It should be emphasized that this lagrangian is gauge
invariant for real or virtual $a_1$ by construction, on the contrary to
the claim made in  Ref.~\cite{xiong}.  
Using this lagrangian, one can derive the amplitude for $a_{1} (k,e)
\rightarrow \pi (p) \gamma (k^{'},e^{'})$ :
\begin{eqnarray}
{\cal M} (a_{1} \rightarrow \pi \gamma) & = &   (T_{2})_{\mu\nu} 
e^{\mu} e^{' \nu} 
\label{eq:t2}
\\
& = & {\kappa_{6} e g f_{\pi} \over
m_{\rho}^2}~\left[ k \cdot k^{'} e \cdot e^{'} -   
k \cdot e^{'} k^{'} \cdot e \right].
\label{eq:ampapig}
\end{eqnarray} 
In Ref.~\cite{ko}, the  condition $k^{'} \cdot e = 0$ has been 
already imposed,  which causes that the amplitude seems to lack in 
gauge invariance.

This amplitude (\ref{eq:ampapig}) predicts that the decay rate
for $a_{1} \rightarrow \pi \gamma$   to be 670 keV for $m_{a_1} =
1260 $ MeV,  which nicely  compares with the experimental data $(640 \pm 
246)$ keV.  We note that other effective lagrangians used in Refs.~
\cite{xiong,song}  predict rather large values of  $\Gamma ( a_{1} 
\rightarrow 
\pi \gamma ) = 1.42 $ MeV.   Thus, the $a_{1}$ contribution to the photon
production rate from hot hadronic matter  made by Xiong {\it et al.}
might have been  overestimated by a factor of $\sim 2$ or so.
This may be also true of dilepton production rate.
In the following, we use the $a_{1} \rho \pi$ vertex given in this section,
and calculate the $a_{1}(1260)$ contributions to photon and the dilepton
 productions from hot hadronic matter.   

\section{Photon production}
\label{sec:photon}

In this section, we consider the $a_{1}$ contribution to photon production 
rate from hot hadronic matter through $\pi \rho \rightarrow a_{1} \rightarrow 
\pi \gamma$ (the $s$ channel exchange only).  
In principle,  $a_{1}$ exchanges in other channels should be included.
However, their effects are small corrections to the $s$ channel 
contribution \cite{song}, and shall be neglected in the following for 
simplicity.

The $a_1$ contribution to  photon production via $\pi (p) \rho 
(k,e) \rightarrow a_{1} \rightarrow \pi (p^{'}) \gamma  
(k^{'},e^{'}) $ is  given by  (Fig.~1)
\begin{equation}
{\cal M} = {e\over g}~g_{a_{1} \rho \pi}~\left[ f_{a_{1} \rho \pi} (s, 
m_{\rho}^{2}) e_{\mu} e_{\nu}^{'} + g_{a_{1} \rho \pi}~
( k + p ) \cdot e ~k_{\mu} e_{\nu}^{'} \right]~
D_{a_1}^{\mu\nu} ((k+p)^{2}, m_{a_1}^{2}),
\end{equation}
where the $D_{a_1}^{\mu\nu} (k^{2}, m_{a_1}^{2})$ is the propagator of 
the intermediate $a_1$ resoance, 
\begin{equation}
 D_{a_1}^{\mu\nu} (k^{2}, m_{a_1}^{2}) = { -i \over  k^{2} - m_{a_1}^{2}}~
 \left( g_{\mu\nu} - { k_{\mu} k_{\nu} \over m_{a_1}^{2} } \right), 
\end{equation}
and the form factors $f_{a_{1}\rho\pi}$ and $g_{a_{1} \rho \pi}$ are given 
by Eqs.~(\ref{eq:amparhopi})--(\ref{eq:garhopi}) in Sec.~\ref{sec:two}.  

The corresponding cross section  can be obtained by integrating the 
following expression over $t$ :
\begin{equation}
{d \sigma_{\gamma} \over dt} (s,t) = {1 \over 64 \pi s p_{c.m.}^2}~
\left| {\cal M} \right|^2.
\end{equation}
The thermal photon  emission rate through  $1 + 2 \rightarrow 3 + \gamma$
can be written  as \footnote{We show the whole expressions in the following, 
although they are lengthy, in order to correct some typos in the Appendix of
Ref. ~\cite{song}.}   
\begin{eqnarray}
E {dR \over d^{3}p} & = & {N\over 16 (2 \pi)^{7} E}~\int_{s_0}^{\infty} ds 
\int_{t_{min}}^{t_{max}} dt ~\left| {\cal M} \right|^{2} ~
\\
&& \int dE_{1} \int dE_{2} ~{f_{1}(E_{1}) f_{2}(E_{2}) \left[ 1 + f_{3}(
E_{1} + E_{2} - E) \right]  \over \sqrt{ a E_{2}^{2} + 2 b E_{2} + c} },
\nonumber   
\end{eqnarray}
where $E$ is the photon energy, $N$ is the number of degeneracy, and   
\begin{eqnarray}
a & = & - (s + t - m_{2}^{2} - m_{3}^{2} )^{2},
\\
b & = & E_{1} ( s + t - m_{2}^{2} - m_{3}^{2})(m_{2}^{2} - t)
\nonumber   \\
& + & E \left[ ( s + t - m_{2}^{2} - m_{3}^{2})(s-m_{1}^{2} - m_{2}^{2}) 
- 2 m_{1}^{2} (m_{2}^{2} - t) \right],
\\
c & = &-E_{1}^{2} (m_{2}^{2} - t)^{2} 
\nonumber  \\
& - & 2 E_{1} E \left[ 2 m_{2}^{2} ( s+t-m_{2}^{2} - m_{3}^{2}) - (m_{2}^{2} 
-t)(s-m_{1}^{2} - m_{2}^{2}) \right]
\nonumber  \\
& - & E^{2} \left[~ (s-m_{1}^{2} - m_{2}^{2})^{2} - 4 m_{1}^{2} m_{2}^{2} 
\right]   
\nonumber  \\
& - & 
(s+t-m_{2}^{2}-m_{3}^{2})(m_{2}^{2}-t)(s-m_{1}^{2}-m_{2}^{2})
\nonumber   \\
& + & m_{2}^{2} (s+t-m_{2}^{2}-m_{3}^{2})^{2} + m_{1}^{2}(m_{2}^{2}-t)^{2}.
\end{eqnarray}
The integration regions for $E_{1}$ and $E_2$ are determined by
\begin{eqnarray}
E_{1} & \geq & {(s+t-m_{2}^{2}-m_{3}^{2}) \over 4 E} + { E m_{1}^{2} \over 
(s + t -m_{2}^{2} - m_{3}^{2}) },
\\
E_{2} & \geq & {(m_{2}^{2}-t) \over 4 E} + {E m_{1}^{2} \over (m_{2}^{2} 
- t)},
\\
E_{2} & \geq & {b\over a} + \sqrt{b^{2} - a c \over a} 
\\
E_{2} & \leq & {b\over a} - \sqrt{b^{2} - a c \over a} 
\\
E_{1} & + & E_{2} - E \geq 0.
\end{eqnarray}
The $s,t$ variables satisfy 
\begin{eqnarray}
s + t - m_{2}^{2} - m_{3}^{2}  \geq 0,
\\
t - m_{2}^{2} \leq 0.
\end{eqnarray}

The resulting photon production rates (in unit of ${\rm fm}^{-4}~{\rm 
GeV}^{-2}$)  are shown in Fig.~\ref{figtwo} for $T = 100, 150 $ and 200 
MeV, respectively, as functions of the photon energy $E_{\gamma} = E$.
  For comparison with other approaches, we show the 
results by Xiong {\it et al.} in the dotted curves, which were obtained
using the approximate analytic formula for the photon production rate 
given in Ref.~\cite{xiong}.    
Our photon spectra have  similar shapes and magnitudes with those 
obtained by Xiong {\it et al.}, although ours are slightly smaller in  
most regions of $E_{\gamma}$.   Thus, the photon production rates from 
the channel $\pi \rho \rightarrow a_{1} \rightarrow \pi \gamma$ in hot 
hadronic 
matters are more or less the same when different effective
lagrangians are employed.  One may conclude that the photon signal from 
quark-gluon plasma is outshined by those from hot hadronic matter  
as a result of the $a_{1}(1260)$ contribution.  This situation is in 
contrast to the $a_{1} (1260)$ contributions to the dilepton signals 
from hot hadronic matter which will be discussed in the following
section.   

\section{Dilepton production}
\label{sec:dilepton}

The dilepton production via $\pi a_{1} \rightarrow \rho \rightarrow l^+ 
l^-$ can be obtained from Eqs.~(\ref{eq:amparhopi})--(\ref{eq:garhopi}) 
with $p \rightarrow -p$ (Fig.~3).  Here, one should be careful that  
there are two contributions to the $a_{1} \pi \rightarrow 
\gamma^*$ vertex, 
one from $\rho-\gamma$ mixing and the other from direct coupling :
\begin{equation}
{\cal M} (\pi a_{1} \rightarrow \rho \rightarrow \gamma^{*} \rightarrow 
l^+ l^- )
\equiv {\cal M}^{\mu}~ {- i \over M^2}~i~\sqrt{4 \pi \alpha}~ 
\bar{v} (l_{1}) \gamma_{\mu} 
u (l_{2}),  
\end{equation}
where the   hadronic vector current ${\cal M}_{\mu}$ is given by
\begin{equation}
{\cal M}_{\mu} = \left[ (T_{1})_{\mu\nu} (-k,-k^{'})~\left(  
{m_{\rho}^{2} \over m_{\rho}^{2} - k^{'2} - i m_{\rho} \Gamma_{\rho} } 
\right)
- (T_{1})_{\mu\nu} (k,k^{'} = 0) \right]~e^{\nu} (a_{1}), 
\end{equation}
where $(T_{1})_{\mu\nu} (k,k^{'}) $ is defined in Eq.~(\ref{eq:t1}). 
The first term  comes from $\rho-\gamma$ mixing, whereas the 
second comes from direct $a_{1} \pi \gamma$ coupling.  Note that the direct  
coupling (the subtracted  term) can be obtained from $(T_{1})_{\mu\nu}$ by  
setting $k^{'} = 0$.  The physical process $a_{1} \rightarrow \pi \gamma$ 
can be obtained from the above amplitude by setting $k^{'2} = 0$ (not 
$k^{'} = 0$), which leads to the same result as Eq.~(\ref{eq:ampapig}), or 
$(T_{2})_{\mu\nu}$. 

One can express $| {\cal M} |^{2}$ by 
\begin{equation}
\left| {\cal M} \right|^{2} = 4 \left( {4 \pi \alpha \over s} \right)^{2} ~
L_{\mu\nu} H^{\mu\nu},
\end{equation}
where  $ s \equiv M_{ll}^{2} = (l_{1}+l_{2})^{2} = (p + k )^2$ and 
\[
L_{\mu\nu} = l_{1\mu} l_{2\nu} + l_{1\nu} l_{2\mu} - g_{\mu\nu} l_{1} 
\cdot l_{2}.
\]
The hadronic tensor $H_{\mu\nu}$ is obtained by replacing 
\[
e_{\mu} (k) e_{\nu}^{*} (k) \longrightarrow 
\left( - g_{\mu\nu} + { k_{\mu} k_{\nu} \over m_{a_1}^2} \right),
\]
in ${\cal M}_{\mu} {\cal M}_{\nu}^{*}$.  
The differential rate for the dilepton production ($1 + 2 \rightarrow 
l^+ l^-$) can be obtained by   the following expression [3] :
\begin{equation}
{dN\over d^{4}x dM_{ll}^{2}} = {\cal N}~{T^{2} \over 2 (2 \pi)^4}~
\sigma (M_{ll}^{2}) 
~\sqrt{M_{ll}^{4} - 2 M_{ll}^{2} ( m_{1}^{2} + m_{2}^{2} ) + (m_{1}^{2} 
- m_{2}^{2}
)^{2} }~G(T,M_{ll}^{2}),
\end{equation}
where ${\cal N}$ is the number of degeneracy, and 
the function $G(T,M^{2})$ is 
\begin{equation}
G(T,M^{2})  =  \int_{m_{1}/T}~dx {1\over e^{x} - 1}~\ln \left( 
{1 - \exp (-y_{+}) \over 1 - \exp (-y_{-}) } \right),
\end{equation}
with
\begin{equation}
y_{\pm}  =  {1 \over 2 m_{1}^2}~\left[ (s - m_{1}^{2} - m_{2}^{2}) x 
\pm \sqrt{ s^{2} - 2 s (m_{1}^{2} + m_{2}^{2} ) + (m_{1}^{2} - 
m_{2}^{2} )^{2} }~\sqrt{x^{2} - {m_{1}^{2} \over T^2} }~~ \right].
\end{equation}

Dilepton production rates from $\pi^+ \pi^-$ 
and $q \bar{q}$ annihilations are given by 
\begin{eqnarray}
\sigma_{q} (M_{ll}^{2}) & = & {4\pi \over 3}~{\alpha^{2} \over M_{ll}^2}~
N_{c} (2 
s + 1)~\Sigma_{f=u,d,s} e_{f}^2, 
\\
\sigma_{\pi} (M_{ll}^{2}) & = & {4\pi \over 3}~{\alpha^{2} \over M_{ll}^2}
~| F_{\pi} (
M_{ll}^{2}) |^{2}~\left( 1 - {4 m_{\pi}^{2} \over M_{ll}^2} \right)^{1/2}.
\end{eqnarray}
For the electromagnetic pion form factor $F_{\pi} (M^{2})$, we have used 
the simple $\rho$ meson propagator with including a higher resonance 
$\rho^{'} (1600)$ as follows \cite{kajantie} :
\begin{equation}
| F_{\pi} (s)|^{2} = {m_{\rho}^{4} \over {(s - m_{\rho}^{2})^{2} + 
m_{\rho}^{2} \Gamma_{\rho}^{2}} } + {1\over 4}~
 {m_{\rho^{'}}^{4} \over {(s - m_{\rho^{'}}^{2})^{2} + 
m_{\rho^{'}}^{2} \Gamma_{\rho^{'}}^{2}} }, 
\end{equation}%.  
with $\Gamma_{\rho^{'}} = 260$ MeV.  
This modified $\rho$ propagator is also used in the $\pi a_{1} \rightarrow
\rho \rightarrow l^{+} l^{-}$.  This may not be possible in any approaches
based on effective lagrangians, but it gives better descriptions. 
Thus we adopt this modification in the following.
This modification of $\rho$ propagator enhances  the dilepton productions
near the $\rho^{'} (1600)$ as expected, but its effects quickly die out
for higher $M_{ll}$ region.  

In Figs.~\ref{figfour} (a),(b) and (c), we show the dilepton production
rates (in unit of ${\rm fm}^{-4}~{\rm GeV}^{-2}$)  from hot
hadronic matter at $T = 100, 150$ and 200 MeV, respectively.
For comparison, dilepton production rates from $\pi^+ \pi^-$
and $q \bar{q}$ annihilations are shown together.  
The solid, the dotted and the dashed curves correspond to the $\pi a_{1}$,
$\pi^+ \pi^-$ and $q \bar{q}$ annihilations into dileptons, respectively.  
Our results show that 
the dilepton signals from the $a_{1}$ contribution via $\pi a_{1}
\rightarrow \rho^{0} \rightarrow l^{+} l^{-}$ is remarkably similar to 
those from the channel $\pi^{+} \pi^{-} \rightarrow \rho^{0} \rightarrow
l^{+} l^{-}$ for $M_{ll} > 1.5$ GeV. Furthermore, both of them are 
negligible compared to the dileptons 
from the quark-antiquark annihilations for high $M_{ll}$ region, 
which is in contrast to the results of Ref.~\cite{gale}.
This difference can be attributed to the different off-shell behaviors of 
the $\pi-a_{1}-\rho (\gamma)$ vertex, {\it i.e.}, to the use of  different 
effective lagrangians. 
Therefore, dilepton signals with $M_{ll} > 2$ GeV is dominated by those
from quark-gluon plasma phase.  

\section{Conclusion}
\label{sec:conclusion}

In this work, we have reconsidered the $a_{1} (1260)$ contributions to 
photon and dilepton signals from hot hadronic matter in the 
framework of the effective lagrangian by Ko and Rudaz, which nicely 
interpolates $a_{1} \rightarrow \rho \pi$ and $a_{1} \rightarrow \pi 
\gamma$ in accordance with the vector meson dominance hypothesis 
\cite{ko}.  

Our result on the  photon production through $\pi \rho \rightarrow 
a_{1} \rightarrow \pi \gamma$ is similar in the shape and the magnitude 
to that obtained by Xiong
{\it et al.} and Song using  different effective lagrangians.
Since different effective lagrangians give similar results, one
may conclude that the photons produced from hot hadronic matter via
$\pi \rho \rightarrow a_{1} \rightarrow \pi \gamma$ overshine those 
from the 
quark-gluon plasma, and photon signals may not be a good candidate for
probing the formation of quark-gluon plasma and its properties.  
 
On the other hand,  the $a_{1}$ contributions to the dilepton production 
rates obtained in this work is smaller than the result 
obtained in Ref.~\cite{gale} by more than an order of magnitude in the 
range of interest, $ m_{\phi} < M_{ll} < 3$ GeV.  Our result is also below 
the dilepton production rate from quark--antiquark  annihilation for high 
$M_{ll}$ region. 
%Thus, in view of these results, it may be premature to discard 
%dilepton signals as a possible probe for the formation of quark--gluon 
%plasma at the early stage of relativistic heavy ion collisions. 
After including dynamics of hadronic matter,  the dilepton 
production from hadronic matter without the $a_1$ resonance is known to be 
more  important than that from qaurk-gluon plasma for 
low $M_{ll}$ region below 1 GeV or so \cite{kajantie}.
Both are comparable in the intermediate range, 
$M_{\phi} < M_{ll} < 2$ GeV, and the latter becomes dominating 
at high $M_{ll}$ region ($M_{ll} > 2$ GeV).
We don't expect this picture would not change substantially even if we
include the $a_1$ resonance along the line considered in this work.
Dilepton signals at high $M_{ll}$ region would be dominated by those
produced at quark-gluon phase.  
Since different effective lagrangians give vastly different predictions
for the dilpeton production rates from hot hadronic matter, one may say 
{\it at least} the dilepton production rates from hot hadronic matter 
are hard to reliably estimate in the language of effective lagrangians 
for hadrons, and are quite model-dependent.
 
\acknowledgements

We are grateful to  Dr. C. Song for useful correspondences.
P.K. was supported in part by  NON DIRECTED RESEARCH FUND 1994, 
Korea Research  Foundations, by KOSEF through CTP at 
Seoul National 
University, and by the Basic Science Research Institute program, 
Ministry of Education, 1995, Project No. BSRI--95--2425.
K.Y.L. and J.K.K. were  supported by KOSEF.
  
% now the references. delete or change fake bibitem. delete next three
%   lines and directly read in your .bbl file if you use bibtex.

% figures follow here
%
% Here is an example of the general form of a figure:
% Fill in the caption in the braces of the \caption{} command. Put the label
% that you will use with \ref{} command in the braces of the \label{} command.
%
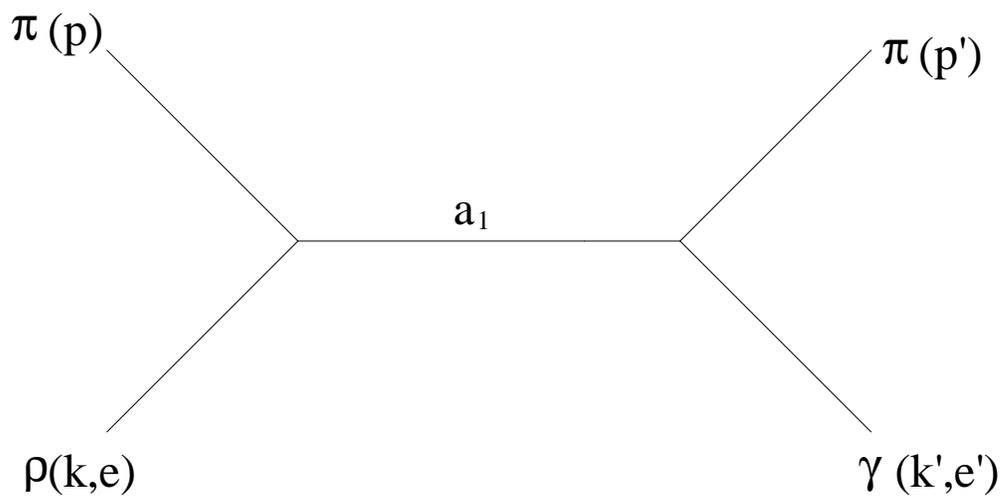
\begin{figure}
\caption{Feynman diagram for $ \pi (p) \rho (k,e) \rightarrow a_{1} 
\rightarrow 
\pi (p^{'}) \gamma (k^{'}, e^{'})$.$~~~~~~~~~~~~~~~~~~~~~~~~~~~~~~~~~~~$}
\label{figone}
\end{figure}

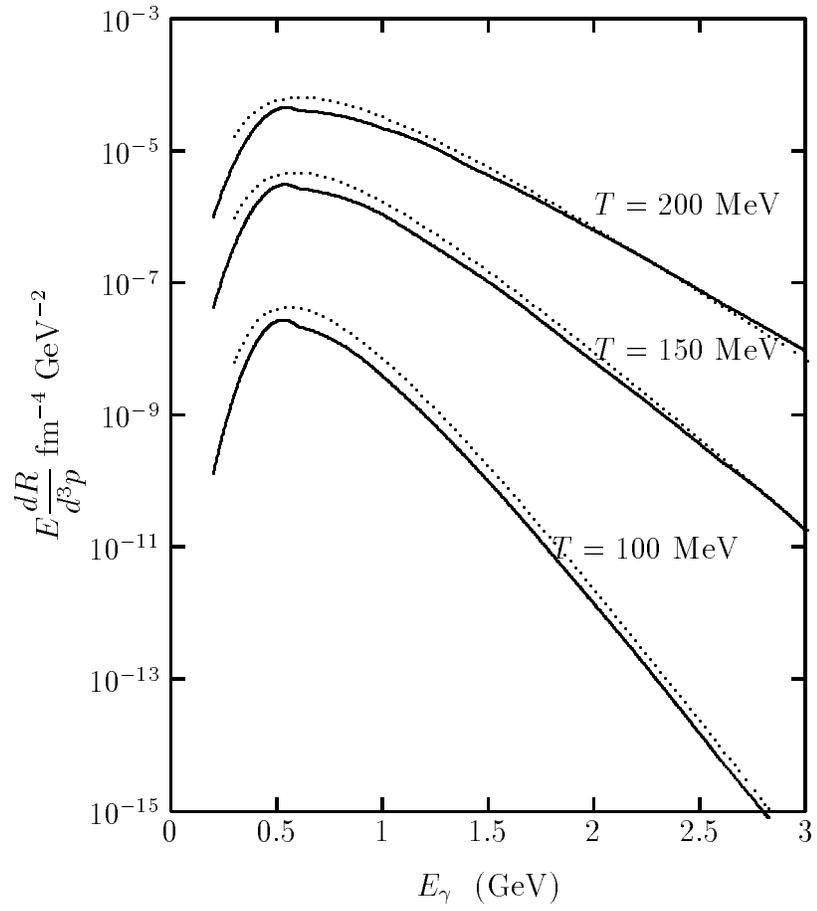
\begin{figure}
\caption{The  photon spectrum from $\pi \rho \rightarrow a_{1} 
\rightarrow \pi \gamma$ for $T = 100, 150$ 
MeV  and 200 MeV. Our results are shown in the solid curves, and those
by Xiong et al. [1] are shown in the dotted curves.
}
\label{figtwo}
\end{figure}

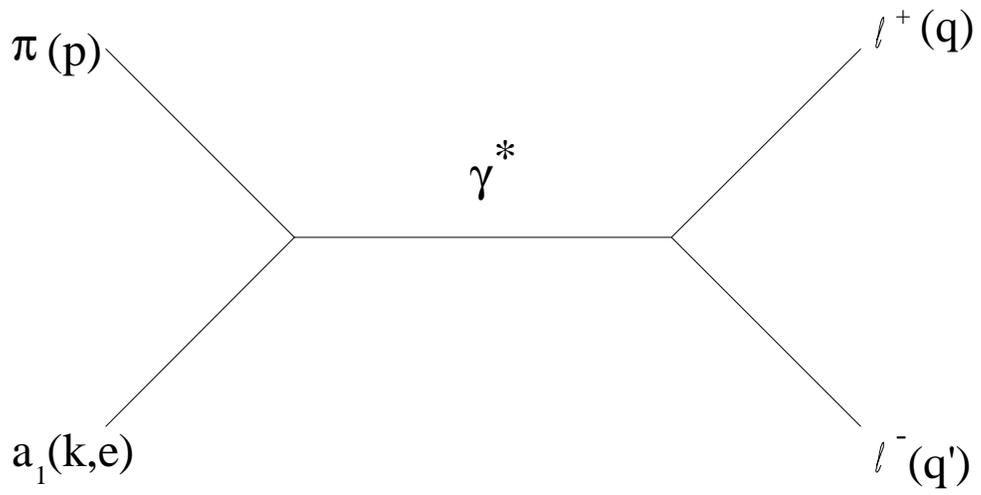
\begin{figure}
\caption{
Feynman diagrams for  $\pi (p) a_{1} (k,e) \rightarrow \rho 
\rightarrow l^{+}(q) l^{-}(q^{'})$ :
contributions from (a) the $\rho-\gamma$ mixing and (b) the direct coupling.
}
\label{figthree}
\end{figure}

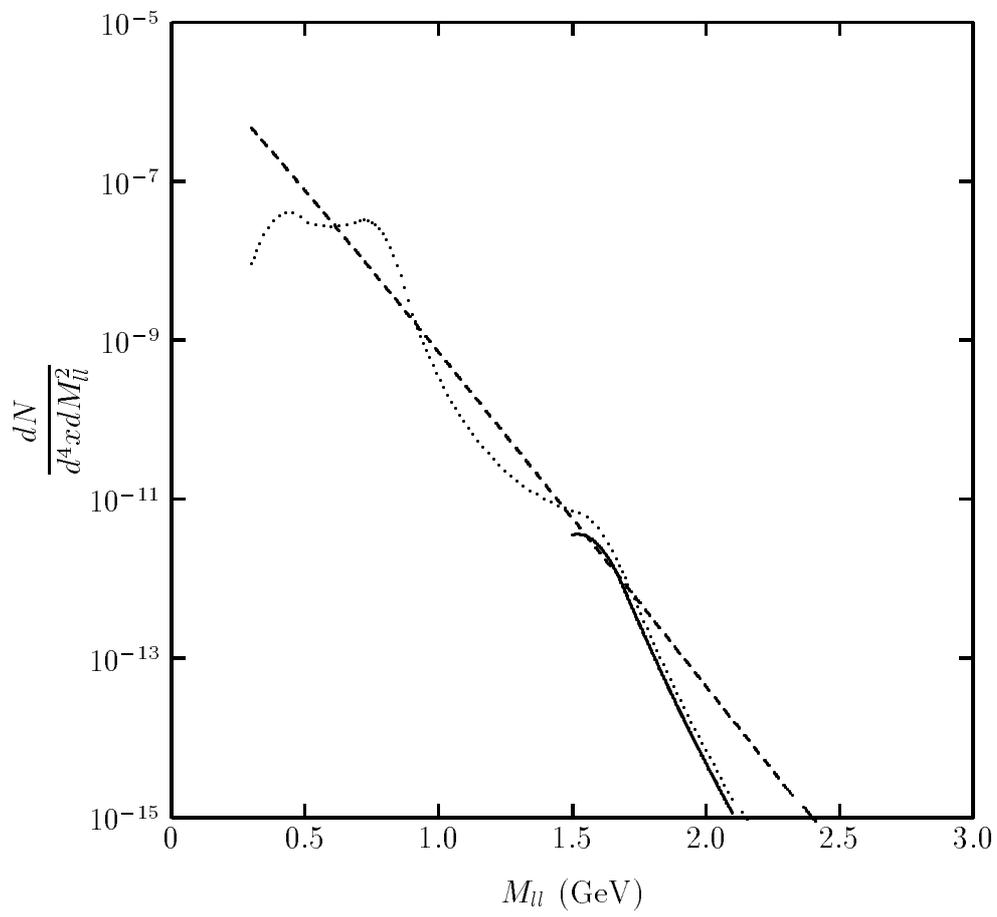
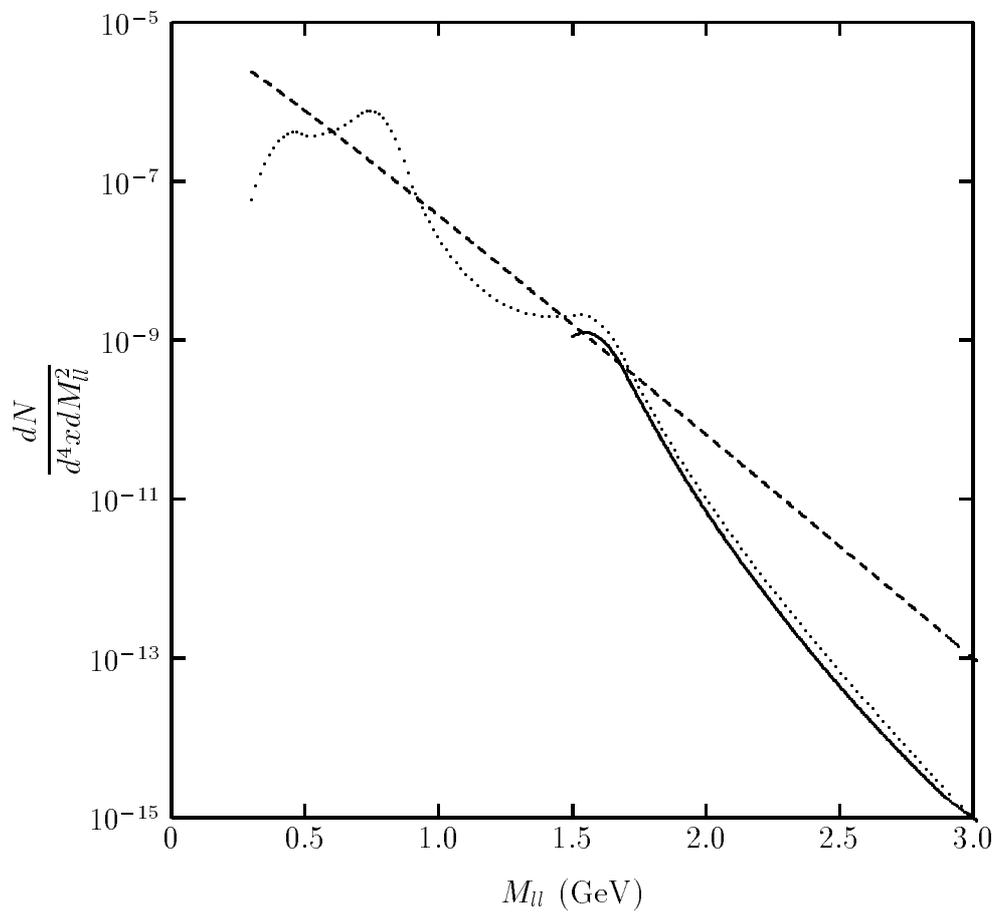
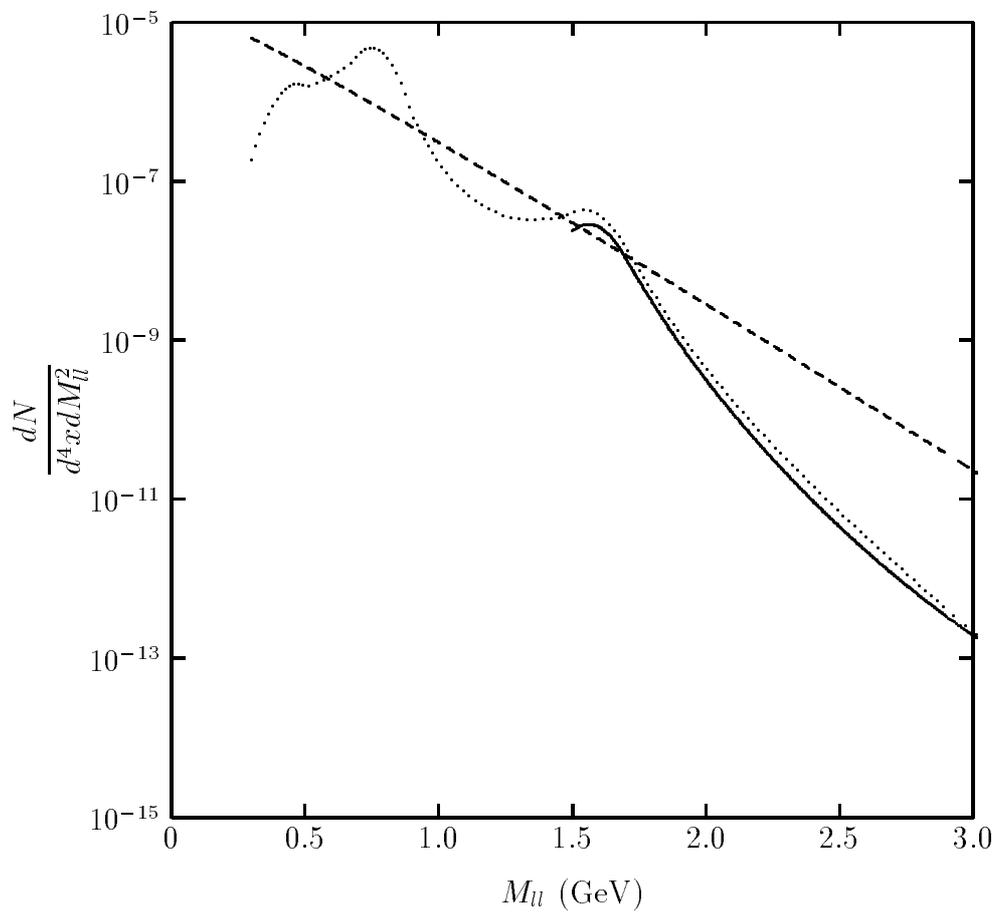
\begin{figure}
\caption{
The dilepton spectrum from $\pi a_{1} \rightarrow \rho  
\rightarrow l^{+} l^{-}$ 
for (a) $T=100$ MeV, (b) $T = 150$ MeV, 
and (c) $T = 200$ MeV in the solid curves.  The dotted and the dashed curves 
represent the contributions from $\pi^+ \pi^-$ and $q \bar{q}$ annihilations 
into dileptons, repectively.
}
\label{figfour}
\end{figure}

% tables follow here
%
% Here is an example of the general form of a table:
% Fill in the caption in the braces of the \caption{} command. Put the label
% that you will use with \ref{} command in the braces of the \label{} command.
% Insert the column specifiers (l, r, c, d, etc.) in the empty braces of the
% \begin{tabular}{} command.
%
% \begin{table}
% \caption{}
% \label{}
% \begin{tabular}{}
% \end{tabular}
% \end{table}

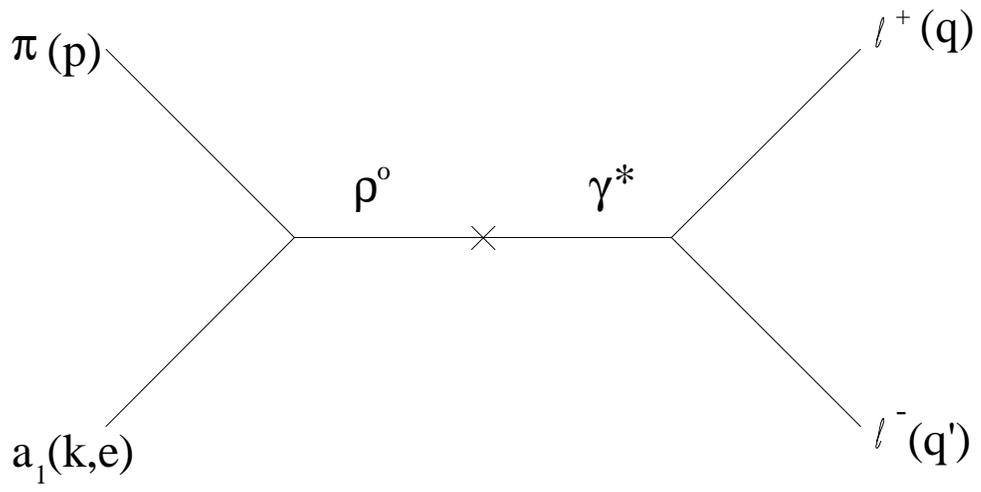
\end{document}